\documentclass{aa}
\usepackage{graphicx}
\usepackage{txfonts}
\begin{document}
   \title{Photometric Survey of the Polar ring galaxy NGC 6822\thanks{Based on 
observations obtained with MegaPrime/MegaCam, a joint
project of CFHT and CEA/DAPNIA, at the 
   Canada-France-Hawaii Telescope (CFHT) which is
 operated by the National Research 
Council (NRC) of Canada, the Institut National des Sciences de l'Univers
of the Centre National de la Recherche Scientifique (CNRS) of France,
and the University of Hawaii.}}
%


   \author{P. Battinelli
          \inst{1}
\and
           S. Demers \inst{2}
\and
	W. E. Kunkel \inst{3}
         } 


   \institute{
INAF, Osservatorio Astronomico di Roma
              Viale del Parco Mellini 84, I-00136 Roma, Italia\\
              \email {battinel@oarhp1.rm.astro.it }
         \and D\'epartement de Physique, Universit\'e de Montr\'eal,
                C.P.6128, Succursale Centre-Ville, Montr\'eal,
                Qc, H3C 3J7, Canada\\
                \email {demers@astro.umontreal.ca }
\and
Las Campanas Observatory, Casilla 601, La Serena, Chile\\
\email  {kunkel@jeito.lco.cl}
}

   \date{Received; accepted}


   \abstract
{We have previously established, from a carbon star survey, that the
Local Group dwarf irregular galaxy NGC 6822 is much larger than
its central bright core.}
{ Four MegaCam fields are acquired 
to survey a 2$^\circ\times$ 2$^\circ$ area centred on 
NGC 6822 to fully determine its extent and map its stellar populations.}
{Photometry of over one million stars is
obtained in the SDSS g$'$, r$'$, i$'$ to three magnitudes below the
TRGB. RGB stars, selected from their magnitudes and colours, are used
to map the NGC 6822 stellar distribution up to a distance of 60 arcmin.}
{We map the reddening over the whole area. 
We establish that the stellar outer structure of NGC 6822 is elliptical 
in shape, with $\epsilon=0.36$ and 
a major-axis PA = 65$^\circ$, contrasting with the orientation of
the HI disk. The density enhancement can be seen up to a semi-major axis
of 36$'$ making NGC 6822 as big as the Small Magellanic Cloud.
We fit two exponentials to the surface density profile of the spheroid,
and identify a bulge with a scale length of 3.85$'$ and an outer spheroid
with a scale length of 10.0$'$. 
We find intermediate-age C stars up to $\sim$ 40$'$ while demonstrating 
that the SDSS filters are unsuitable to identify extragalactic C stars.}
{NGC 6822 is a unique Local Group 
galaxy with shape and structure suggesting a polar ring
configuration. Radial velocities of carbon stars have indeed
demonstrated that there are two kinematical systems in NGC 6822.}



\maketitle
%

\section{Introduction}
Recent carbon star surveys of Local Group dwarf irregular galaxies (dIrrs)
(Letarte et al. 2002; Demers et al. 2004) have demonstrated that
some dIrrs are much bigger than assumed from their bright central
core. Indeed, the NGC 6822 giant branch is seen up to the edge of the
CFHT12k 42$' \times 28'$ field. Do these extended structures correspond
to a huge stellar halo or do we see an extended disk, as reported by Gallart
et al. (2004) for the Large Magellanic Cloud? However, in the presence
of evidence for disruptive processes on their periphery, galaxies with recent
harassment episodes admit a variety of interpretations of what formation
mechanisms acted at the peripheries. The LMC is quite big,
its stellar populations are spread over some  20 degrees
(Irwin 1991; van der Marel 2002), being so close to the Milky Way its
structure must have been influenced by the later. 

NGC 6822 is also known as Barnard's galaxy because he discovered it 
(Barnard 1884) and described it as a 2$'$ nebula, diffuse and difficult
to see with his 5-inch instrument. NGC 6822
is located in the constellation Sagittarius in the direction of the 
Galactic centre ($\ell$ = 25$^\circ$, b = --18$^\circ$). It is thus
seen behind a relatively heavy stellar foreground and its reddening is
far from negligible. A distance of (m -- M)$_0$ = 23.46 $\pm$ 0.10 
has been obtained from the magnitude
of the tip of the red giant branch by Lee et al. (1993). However, recent
estimates bring the galaxy slightly closer. 
Pietrzy\'nski et al. (2004) derived 
(m -- M)$_0$ = 23.34 $\pm$ 0.05 from  V,I observations of over one hundred
Cepheids. The observations of RR Lyrae by Clementini
et al. (2003) yields a distance of (m -- M)$_0$ = 23.36 $\pm$ 0.17.
Near infrared observations of the tip of the red giant branch by Cioni \&
Habing (2005) give (m -- M)$_0$ = 23.34 $\pm$ 0.12.
We adopt a weighted mean of (m -- M)$_0$ = 23.35. At 470 kpc, NGC 6822
is the nearest dIrr after the Magellanic Clouds. It is isolated in the
Local Group, without close neighbours. 
  
NGC 6822 is one of the few dwarf galaxies surrounded by a huge HI
envelope which was first mapped by Roberts (1972) to a density of 
9 $\times$ 10$^{19}$ cm$^{-2}$. The HI component, recently surveyed by
de Blok \& Walter (2000a, 2000b), to a sensitivity of 1.6 $\times$ 10$^{19}$
cm $^{-2}$, is essentially a disk whose high-resolution
rotation curve has been obtained by Weldrake et al. (2003).  A photometric
survey by Komiyama et al. (2003) revealed the presence
of a significant number of young stars throughout the HI disk.

The LMC, a more luminous galaxy,
seen at the distance of NGC 6822 would subtend
an angular diameter of 2.2$^\circ$.
It is then reasonable to expect that the overall diameter of NGC 6822
could reach one degree. 
Since the CFH12k was found to be inadequate to properly establish the
extent of NGC 6822, we decided to take advantage of
the newly CFHT  MegaCam camera to survey the outer
parts of NGC 6822. 

In Sect. 2 observations are presented. The absorption map over the whole 
observed region along with the reddening free colour-magnitude and 
colour-colour diagrams are presented in Sect. 3. The stellar density profile
of NGC 6822 is obtained and discussed in Sect. 4.   

\section{Observations}

Our observations consist of images taken with MegaPrime/Megacam in Queue 
Observing mode in May and June 2004.
The wide field imager Megacam consists of 36 2048 $\times$ 4612 pixel
CCDs, covering nearly a full 1$^\circ \times 1^\circ$ field. It is mounted
at the prime focus of the 3.66~m Canada-France-Hawaii Telescope.
It offers a  resolution of 0.187 arcsecond per pixel.
Four slightly overlapping fields were observed, the galaxy located
at the common corners, to cover essentially a 2$^\circ \times 2^\circ$ area
centred on NGC 6822. Images were obtained through  g$'$, r$'$ and i$'$.
These filters are designed to match 
Sloan Digital Sky Survey (SDSS) filters, as defined by the Smith et al.
(2002) standards.
Exposures times range from 160 s for i$'$ to 1200 s for
g$'$ to reach i$'$ = 22, r$'$ = 23, g$'$ = 24
 with a S/N = 20.
The observations were secured under excellent seeing.
Two additional Megacam fields, North East  and South West
of the four central fields,
were obtained to better assess the 
foreground stellar surface density. 

The data distributed by the CFHT have been detrended. This means that the
images have already been corrected with the master darks, biases, and
flats.  This pre-analysis produces 36 CCD images, of a given 
mosaic,  with the same zero point and magnitude scale. 
The photometric reductions were done by Terapix, the data reduction
centre dedicated to the processing of extremely large data flow. The Terapix
team, located at the Institut d'Astrophysique de Paris, matches and stacks all
images taken with the same filter and, using SExtractor (Bertin \&
Arnouts 1996), provides magnitude calibrated
catalogues of objects in each of the combined images. SExtrator
classifies objects into star or galaxy but the classification scheme breaks 
down for faint magnitudes, thus not useful in our case.
Furthermore,  a flag is attached to each object: flag = 0 corresponds to 
isolated object not affected by neighbours. As expected, 
few stars in the central part of NGC 6822 have a flag = 0. In this paper
we do not take into account the flags and select the stars according to
their photometric errors.

Because the astrometric
calibration of the images has been done by the CFHT Service Observing
team,
we have equatorial coordinates as well as calibrated colours and
magnitude for each object in the whole field. The six Megacam fields
have been matched and concatenated into three files corresponding to 
each filter. These files contain approximately one million star each.
Many stars have, however, 
large photometric errors. When the error limit is set to $\sigma_{mag} < 0.10$,
the numbers in each file are reduced to $\sim$ 650,000.
Cross identification of the i$'$ and r$'$ file, with a colour requirement
of $\sigma_{ri} < 0.15$ leads to some 650,000 stars while for the three
colour file, with $\sigma_{irg} < 0.125$ we have $\sim$ 500,000. The 
concatenation of the three files reduces even more the number of
stars satisfying these criteria.

   \begin{figure}
   \centering
\includegraphics[width=10cm]{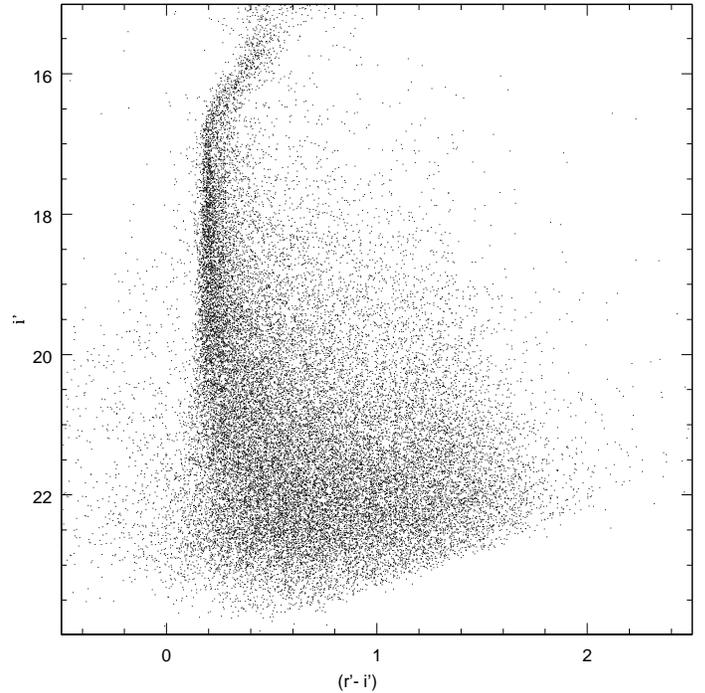}
   \caption{Colour-magnitude diagram of part of the SW foreground field,
30000 stars are plotted               }
              \label{FigCMD}
    \end{figure}

\section{Results}

\subsection{The absorption map}

For a wide and relatively low latitude field, a certain variation of
the foreground reddening is expected. It is important to take into account
differences in the local absorption since they effect directly the number of
 star detected locally. 
To determine the reddening variations across our field, we adopt
a procedure similar to that recently employed with success for IC 10
by Demers et al. (2004).
The basic idea is to compare the observed colour of
the vertical ridge, well visible in the CMD shown in Figure 1 
at $(r' - i')_{ridge}\approx 0.3$, with its unreddened value.
 In this figure we display a representative colour-magnitude diagram (CMD)
 of the foreground toward NGC 6822. This CMD contains 30 000 stars and 
corresponds to a 59$'$ $\times$ 12$'$ region in the 
SW foreground field. Only stars with colour error, as given by SExtractor,
$\sigma_{ri}$ $<$ 0.20 are plotted. Inspection of this CMD is instructive.
Stars brighter than i$'$ $\approx$ 16.5 are saturated; the luminosity
function turnoff is at i$'$ $\approx$ 21.5. 
The well-defined vertical
ridge corresponds to the Galactic G dwarfs, seen along the line of sight,
and located at their main sequence turnoff. 
The unreddened value for the ridge colour,
$(r'-i')_{0,ridge} = 0.12$,
comes from Smith et al. (2002) transformation of R--I into
(r$'$--i$'$) and the G dwarf turnoff at (R--I)$_0$ = 0.35 (Cox 2000).
As a first step, in the CMD we select only the stars in a vertical
strip 0.3 mag wide in colour, centred  on the ridge. We select exclusively
in the 17.0 $<$ i$'$ $<$ 18.5 magnitude
interval in order to have a sample mostly consisting of foreground stars
because fainter objects would include many unresolved galaxies. 
The faintest G dwarfs should then be $\sim$ 3 kpc above the plane of the 
Milky Way. 

The whole field is then subdivided into grid points separated by 100 pix 
and at each  such locus the ridge colour is determined within 2000 pix
radius centred on the grid intersection. The local
reddening value is thus given by $E(r'-i') = (r' - i')_{ridge}-(r'-i')_{0,ridge}$.
We observe a range in colour for the ridge between 0.24 and 0.32
corresponding to reddening ranging from 0.12 $< E(r'-i')< 0.20$ , while
the  extinction in the $i'$ band, 0.37$< A_{i'} < 0.64$.
The standard colour excess, E(B--V) varies from 0.19 to 0.30.
The global reddening value of E(B--V) = 0.25 adopted by van den Bergh (2000)
appears quite reasonable. Values as high as E(B--V) = 0.45 were determined in
the NGC 6822 centre, from UBV photometry, by Massey et al. (1995).  
Our reddening is the foreground reddening and certainly
does not apply to the inner parts of NGC 6822.

Figure 2 shows the grey-scale foreground extinction towards NGC 6822
along with the iso-absorption contours corresponding to $A_
{i'}=0.45,0.50,0.55$. 
The reddening in this six degree region is patchy and certainly not negligible.
Magnitudes and colours of individual stars are corrected 
using
the absorption map in Fig. 2 and the following relations by Schlegel et
al.
(1998): $A_{i'}=3.1606\times E(r'-i')$ and $E(g'-r')=1.576\times E(r'-
i')$.

   \begin{figure}
   \centering
\includegraphics[width=9cm]{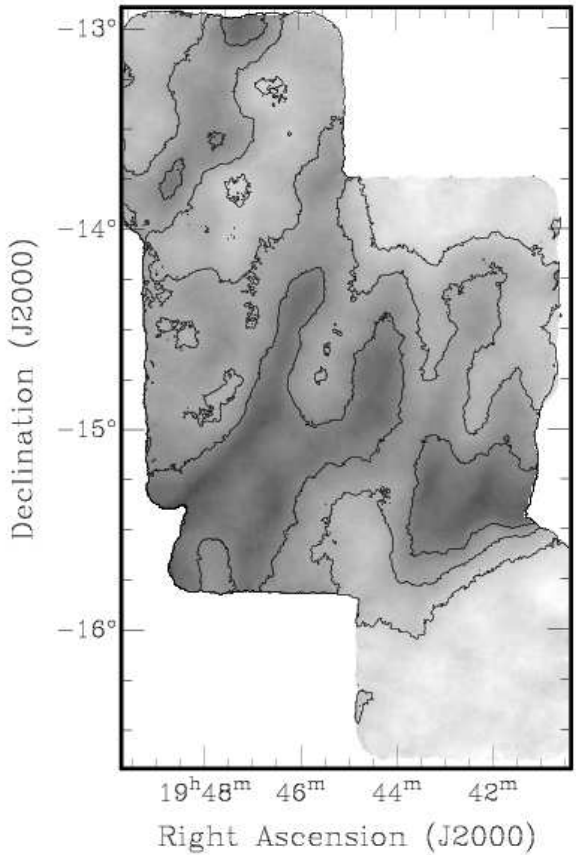}
   \caption{Extinction map of the field surrounding NGC 6822.
Contours correspond to $A_ {i'}=0.45,0.50,0.55$.
               }
              \label{Fig redd}
    \end{figure}

\subsection{Colour-magnitude diagram}
The study of the stellar populations in NGC 6822 is outside the goal of the 
present paper which is focused on the size and shape of this galaxy. For this
reason the CMD is used here only for the selection of a suited subsample of 
stars to be employed for star counts in Sect. 3.4.
   \begin{figure}
   \centering
\includegraphics[ width=9cm]{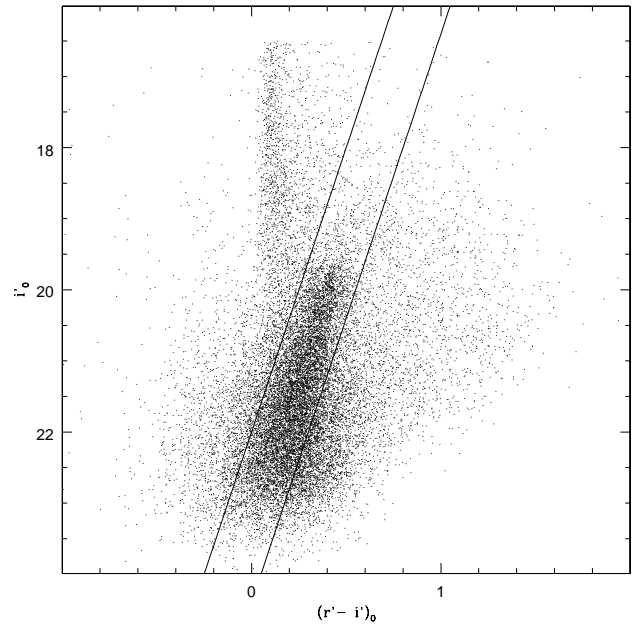}
   \caption{Representative CMD of the central part of NGC 6822. Stars
between the lines are considered as RGB stars.}
              \label{Fig cmd giants}
    \end{figure}
In Figure 3 the reddening corrected CMD of NGC 6822 is shown. The RGB is well 
visible extending for more than 3 magnitudes. 
The surface density of foreground stars in the direction
of NGC 6822 is quite high and therefore,
to detect the low density outskirts of NGC 6822, it is recommended to increase
the contrast as much as possible by selecting stars expected to belong
to the galaxy. The most straightforward method is to consider only stars with
magnitudes and (r$'$ - i$'$) colours corresponding to the red giant
branch of NGC 6822. To do so, we selected stars in the unreddened CMD which
fall between the two lines traced on Fig. 3. 

\subsection{Colour-colour diagram}
NGC 6822, being a Magellanic-type galaxy, young blue stars are
expected to be present. They are indeed present in the central
part where star formation is still taking place.
(Gallart et al. 1996,
Battinelli et al. 2003). The young stars are seen to follow
more or less the HI disk. 
The fainter  outer
region contains almost exclusively older stars. The 
reddening-corrected
colour-colour diagrams, presented in Figure 4, confirm this 
known fact. We display stars with $\sigma_{irg}$ $<$ 0.125
in various radial zones, centred on NGC 6822.  It is our standard
procedure to display such diagrams with only stars with small photometric
errors in order to better see their features. The zones are selected
to contain 25,000 to 30,000 stars.
   \begin{figure*}
   \centering
\includegraphics[ width=9cm]{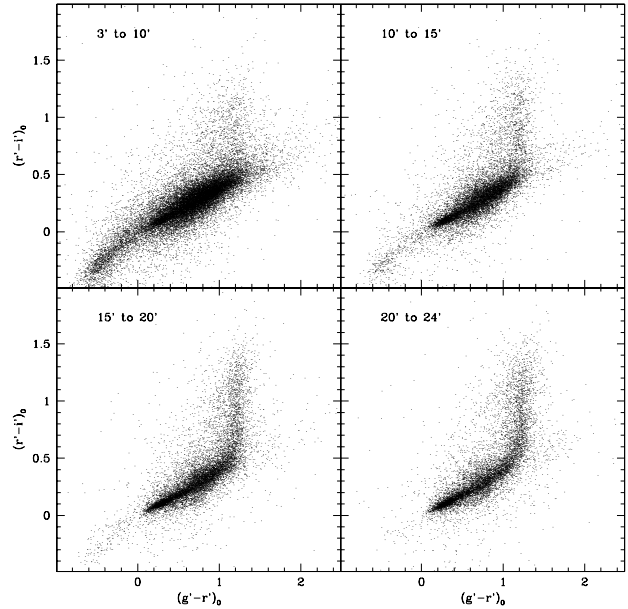}
   \caption{The reddening corrected colour-colour diagrams for four
annuli in NGC 6822.
}
              \label{Fig cc4}
    \end{figure*}

One sees in panel a) the blue main sequence stars, possibly as early
as O5 (see main sequence colours from Fukugita et al. 1996) extending
to negative colours. The bulk of stars are earlier than mid K-type.
In panel b) the vertical branch, corresponding to M-type is stronger
while the very early types have diminished. As one moves further out
the number of blue stars drastically decreases while the
importance  of M-type stars increases. In all the panels, stars
with (g$'$--r$'$)$_0$ $>$ 1.5 and (r$'$--i$'$)$_0$ $\sim$ 0.5 are
seen. These stars have colours corresponding to C stars 
(Margon et al. 2002, Downes et al. 2004). Because of the heavy
foreground pollution it is advisable to have a good estimate of
its influence on the interpretation of the features of
 the colour-colour diagram.
   \begin{figure}
   \centering
\includegraphics[ width=8cm]{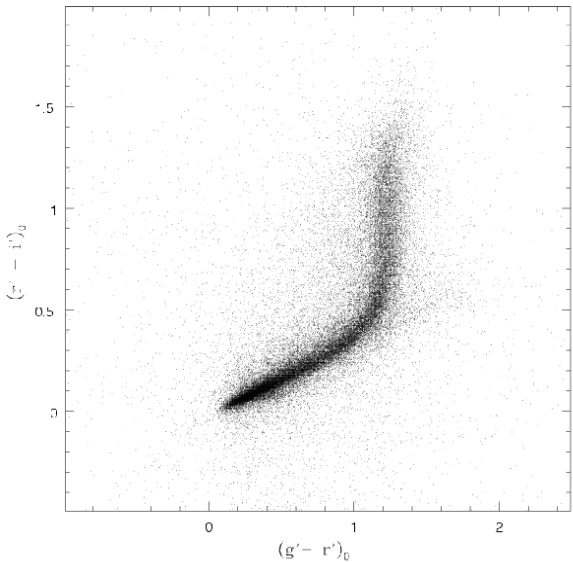}
   \caption{The reddening corrected colour-colour diagram of the 
SW foreground field. 
}
              \label{Fig CCfg}
    \end{figure}

We present, in Figure 5,
a reddening free colour-colour diagram of the one square degree south
west foreground field. 
The main sequence branch, extending from late A-type
($\sim$ 0,0) to late M-type, is well visible in Fig. 5. There is also present,
as a well-defined red tail extending to nearly $(g'-r')_0$ = 1.8, the region
 where carbon stars are located.  The detection of carbon stars in the 
outer regions of NGC 6822 would be of great interest since they are excellent 
kinematic probes.
 Unfortunately, since we do not expect to see the NGC 6822 C star
population extending so far from the galaxy, Fig. 5 suggests that the tail 
is not exclusively populated by C stars and must contain also foreground and/or 
background objects. 

To try to answer this question we follow the procedure used by
Demers \& Battinelli (2005) when they studied the SLOAN colours of
the M31 C star population. We cross-identify the 900 known C stars of
NGC 6822, selected from R, I, CN and TiO photometry, with our reddening
corrected database. Excluding the crowded $10'$ $\times$ $10'$ centre
and taking only stars whose coordinates match within one arcsec ($\sim$ 500 stars),
we obtain 
the colour-colour diagram displayed in Figure 6.  Even though the 
previous sample corresponds to cool N-type C stars with (R--I)$_0$
$>$ 0.90 (Letarte et al. 2002), their (g$'$ -- r$'$) colours span over a
wide range. C stars with (g$'$ -- r$'$)$_0$ less
than $\approx$ 1.6 cannot be differentiated from the bulk of K stars.
The parallel lines, drawn by eye, limit
the region where the bulk of the C stars are found. The luminosity
function of the matched stars is shown in Figure 7. As expected the
magnitude distribution of C stars follows a narrow Gaussian (see e.g. Battinelli \&
Demers, 2005).
  \begin{figure}
   \centering
\includegraphics[ width=6cm]{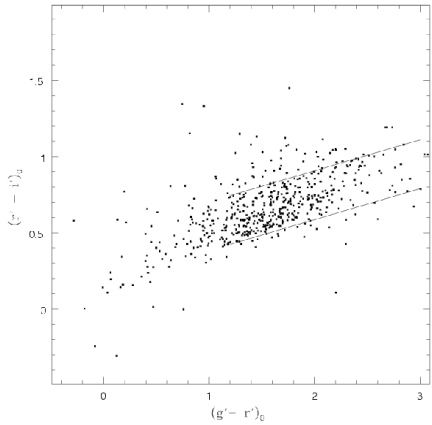}
   \caption{The colour-colour diagram of the C stars cross-identified
with the known C stars. The parallel lines limit the C star region.
}
              \label{Fig C cc plot}
    \end{figure}

   \begin{figure}
   \centering
\includegraphics[ width=6cm]{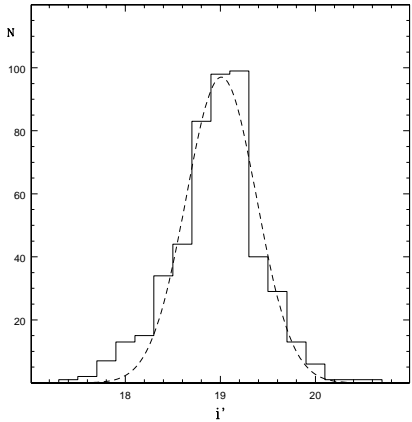}
   \caption{The luminosity function of the C stars of Fig. 6,
we see the usual Gaussian distribution with a $\sigma$ = 0.35
}
              \label{Fig C LF}
    \end{figure}

We then selected the ``C star'' candidates in the two foreground fields
as stars with (g$'$ -- r$'$)$_0$ $>$ 1.5 and having colours within the
two lines drawn on Fig. 6.
Figure 8 presents the LF of these two samples. The Gaussian LF of
NGC 6822 C stars in outlined in each panel, there is little evidence
for a peak at $i'$ $\sim$ 19. We conclude, as expected,  that these stars 
are not C stars belonging to NGC 6822.

The possibility that we are observing numerous dwarf C stars
along the line of sight must be rejected.
Indeed, the surface density of dwarf C stars
is quite low at high Galactic latitude, values such as 0.02 per deg$^2$
are quoted (Margon et al. 2002). Their numbers in the disk, thick disk or bulge are
not well known. Taking into account the origin of dwarf C stars,
de Kool \& Green (1995) estimate that their density in the Galactic disk is
$\sim$ 1$\times$ 10$^{-6}$ pc$^{-3}$. The absolute magnitude of dwarf
C stars is believed to be M$_V$ $\sim$ +10 (Harris et al. 1998). For
such a luminosity, stars at 2 kpc from the Sun 
will already be fainter than i$'$ = 20,
thus the observed number of stars with the right colours
 is much too high to be explained by dwarf C stars. 

Quasars, with z $>$ 3,
 partially overlap the C star region in the colour-colour plane 
(Fan 1999; Fan et al 2001, Richards et al 2001)
since their numbers increase quite a bit at $i'\approx$ 20, they 
could account for some of the interlopers. They would not, however,
define a narrow red tail as the one seen on Fig. 5. On the other end,
galaxies, at the r$' \approx$ 19 level, are found nearly exactly in
the C star region, as Fig. 4 of Newberg et al. (1999) clearly
demonstrates. 
We then conclude that faint galaxies pollute the C star region of the
colour-colour diagram in such a way to render the SDSS colours of little
use to detect C stars. We do see, however, a recognisable difference
between the red tail of Fig. 5 and Fig. 6: C stars often have very red
(g$'$--r$'$). For the subsample of C stars with (g$'$--r$'$) $>$ 1.7, 37\%
of them are redder than (g$'$--r$'$) = 2.0, while only 10\% of the objects
of the red tail of Fig. 5 have such red colours. Thus, objects with
extreme (g$'$--r$'$) have more chances of being C stars than background
galaxies. In the two foreground fields we observe in the C star region of the 
colour-colour diagram (as defined above) 230 interlopers per deg$^2$ for 
i$'$ $<$ 20.0.

   \begin{figure}
   \centering
\includegraphics[ width=6cm]{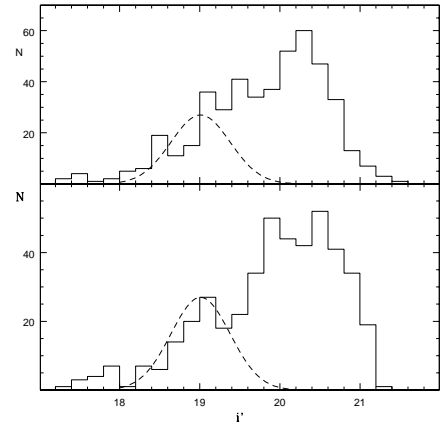}
   \caption{The luminosity function of stars in the NE and SW foreground
fields having colours matching C stars. The Gaussians are reproduced
and scaled from Fig. 7. 
}
              \label{Fig C LF}
    \end{figure}

\subsection{Star counts and the shape of NGC 6822}

Nearly 200,000 stars fall into the RGB region defined in Sect. 3.2. Their 
spatial distribution is shown in Figure 9. 
The density enhancement, at the centre, corresponds
to NGC 6822. Other features seen in Fig. 9 require comments. There is 
an obvious deficiency of stars are the very centre of NGC 6822 which is
explained by crowding and by the inadequacy of SExtractor
to deal with crowded fields. The horizontal empty strips  correspond to the
gaps ($\approx 80''$
wide) between the CCDs.
This complicate the determination of the surface density. 
The narrower vertical gaps (13$''$) are more numerous but not as 
critical. The dent
seen on the lower right of the MegaCam field is due to the failure of three
CCDs during the observation of the SW field. 
Finally, lacks of stars, here and there are due to very 
bright stars contaminating their surrounding.

In order to determine the stellar surface density the gaps need to be
filled. The horizontal gaps are filled by duplicating the observe stellar
distribution on a strip of width 40$''$ above and 40$''$ below the gap.
Similar fill-up is done for the vertical gaps.
The whole field is then covered by a 50 $\times$ 50 pixel wide grid 
and stars are counted  over a circular 500 pixel sampling area, centred
on each intersection of the grid. This is done to  smooth out major 
irregularities (mainly due to bright foreground stars that locally prevent 
the detection of fainter members of NGC6822).

   \begin{figure}
   \centering
\includegraphics[ width=9cm]{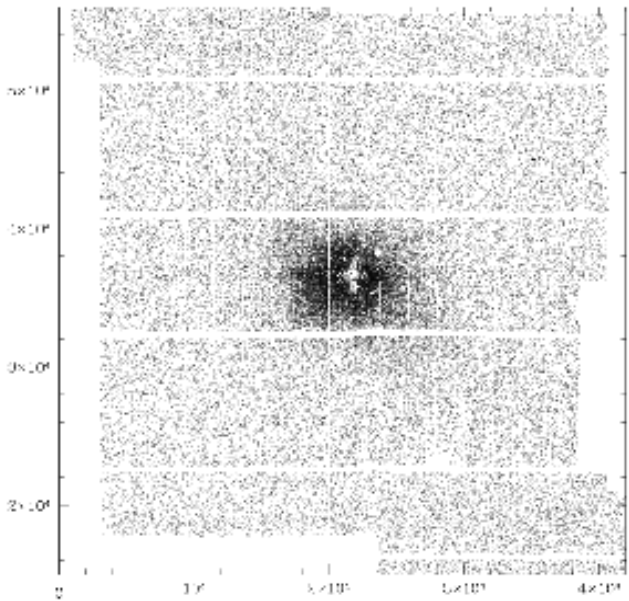}
   \caption{Spatial distribution of the stars located in the RGB region.
North is up and East to the left. The field is about 2$^\circ \times
2^\circ$. Coordinates are in pixels. Only parts of the NE and SW foreground
fields are seen here.
}
              \label{Fig map RGB}
    \end{figure}
The density map is then
transformed into a density image that is analysed with IRAF/STSDAS/ELLIPSE
task to fit isodensity ellipses, determine their position angles and
ellipticities. The results of this task are summarised in Figure 10 where
we show the position angles (PA) and the ellipticities ($\epsilon$) derived
for 
semi-major axes
 larger than 10$'$. Closer to the centre we feel
that our data are inadequate  to properly investigate the isodensity contours.
Furthermore, the stellar distribution in the centre is highly spotty
and asymmetric. Fig. 10 shows that the PA's, close to 80$^\circ$ near
the centre, decrease slightly to settle at $\sim 65^\circ$ at
semi-major axis
larger than 25 arcmin. The isodensity contours become somewhat more 
elliptical as the ellipses increase in size. 
   \begin{figure}
   \centering
\includegraphics[ width=9cm]{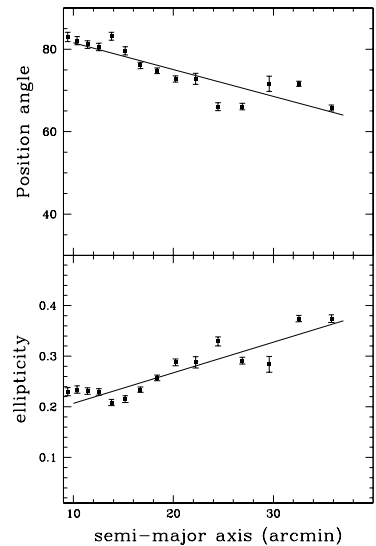}
   \caption{The variation of the PA (upper panel) and the ellipticity (lower
panel) as a function of the major axis of the isodensity elliptical contours.
The lines are a least-square fit.
}
              \label{Fig PA ell}
    \end{figure}
The outer ellipse that can be detected above the noise has
  a semi-major axis of 36$'$. This ellipse is traced in 
Figure 11 over a Digitized Sky Survey image of NGC 6822. The HI isodensity
contours from de Blok \& Walter (2000b) are also traced. This figure
reveals that the stellar spheroid, surrounding the bright central bar of 
NGC 6822 is comparable in size to the HI disk. 
   \begin{figure*}
   \centering
\includegraphics[ width=9cm]{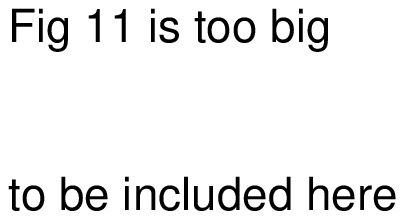}
   \caption{An image of NGC6822 with the HI disk. The ellipse represents
the larger isodensity contour that can be seen. The field is about $2^\circ
\times 2^\circ$.
}
              \label{Fig Ellipse HI}
    \end{figure*}

Using the above PA's and ellipticities, we determine the stellar surface
density profile by counting ``giant'' stars in elliptical annuli. The profile
is shown in Figure 12. As expected, the incompleteness for r $<$ 5$'$ is
severe. The surplus of stars, over the foreground, is seen at least up
to $\approx$ 40$'$. This demonstrates that NGC 6822 is indeed quite large.
We adopt for the foreground contribution, to be remove from the total
counts, a density of 6.93 $\pm$ 0.14 ``giants'' arcmin$^{-1}$, this value
represents the mean of counts in areas with  r $>$ 50$'$.

   \begin{figure}
   \centering
\includegraphics[ width=9cm]{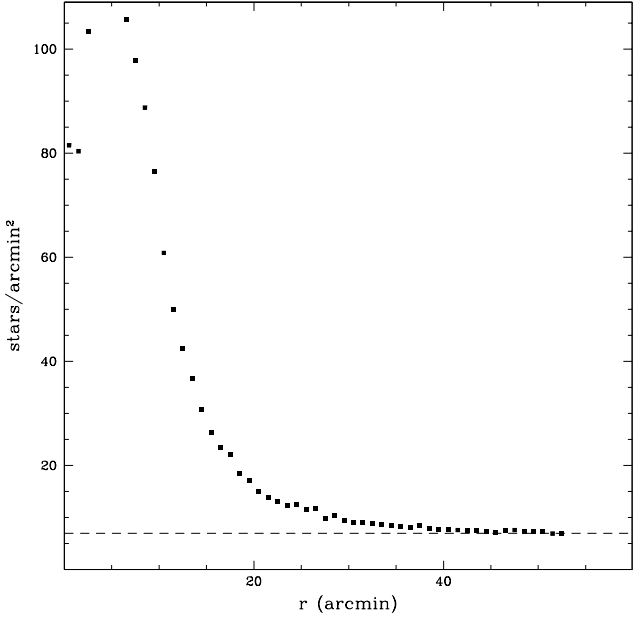}
   \caption{Surface density profile of the stars with magnitudes and colours 
corresponding
to red giants (see Sect. 3.2). The over density, above the foreground represented by the
dashed line, is seen up to 40 arcmin.
}
              \label{Fig Profile}
    \end{figure}
\subsection{The structure of NGC 6822}
As a first step, we fit 
a simple exponential to the profile displayed in Fig. 12,
with three free parameters of the form:
$$ \rho = \rho_{fg} + \rho_oe^{-r/h},\eqno(1)$$
we fit the density in the 7$'$ to 40$'$ interval and obtain:
$\rho_{fg}$ = 7.85 $\pm$ 0.32, $\rho_o$ = 402.2 $\pm$ 14.5 and a
scale length h = 5.16 $\pm$ 0.11$'$. The resulting foreground density
is, however, too large being more than 6 $\sigma$ above our 
estimated value. 
We must
conclude that a single exponential cannot represent the entire
profile of the spheroid. 

It has been the norm, in recent years, to fit the surface brightness 
profile of dE's with a S\'ersic law (i.e. Durrell 1997, Aguerri et al. 2005).
It is therefore reasonable to try a similar fit for our observed density profile.
The S\'ersic (1968) law, given in the form:
$$ \rho = \rho_oe^{{-(r/b)}^n},\eqno(2)$$
is more flexible than the simple exponential 
because it includes a third free parameter ($n$) 
which controls the shape of the profile. This is normally applied 
to the surface brightness but can as well be used for the stellar density.

The fit yields: $\rho_o$ = 3323, $b$ = 0.711 and $n$ = 0.54. This last
coefficient is function, with appreciable scatter,  of the total magnitude 
of the galaxy.  When compared to early-type dwarfs of the Virgo Cluster
(Binggeli \& Jerjen 1998), NGC 6822, with B$_T$ = --15.2,
should have 0.5 $< n <$ 1.0. At the distance of NGC 6822, the scale
length of 0.711 arcmin corresponds to 97 pc.
The S\'ersic laws implies that the density decline is shallower than a simple
exponential. 
It turns out that the solution of the S\'ersic fit is very much
dependent on the radial interval selected and $n$ tends toward 1 when we
limit the maximum r to 40$'$ or smaller values, 
suggesting that the inner spheroid can 
indeed be represented by a single exponential.

Having the density profile over a long interval provides an 
opportunity to see if a two-exponential law could also fit the observed density
variation. Indeed, we can obtain a slightly better fit with two
exponentials than with the S\'ersic law. Fitting the whole 10$'$ to
50$'$ interval and using our determine foreground density, 
we obtain:

$$ \rho = Ae^{-r/a} + Be^{-r/b},\eqno(3)$$

with A = 45.2 $\pm$ 16.8, a = 10.03 $\pm$ 1.12$'$, B = 579.1 $\pm$ 55.0
 and b = 3.85 $\pm$ 0.26$'$ for a correlation 
coefficient r = 0.999 (see Figure 13). With a short scale length of 560 pc, NGC 6822
is quite normal for its luminosity when compared to the dIrr sample of
Padori et al. (2002)

   \begin{figure}
   \centering
\includegraphics[ width=9cm]{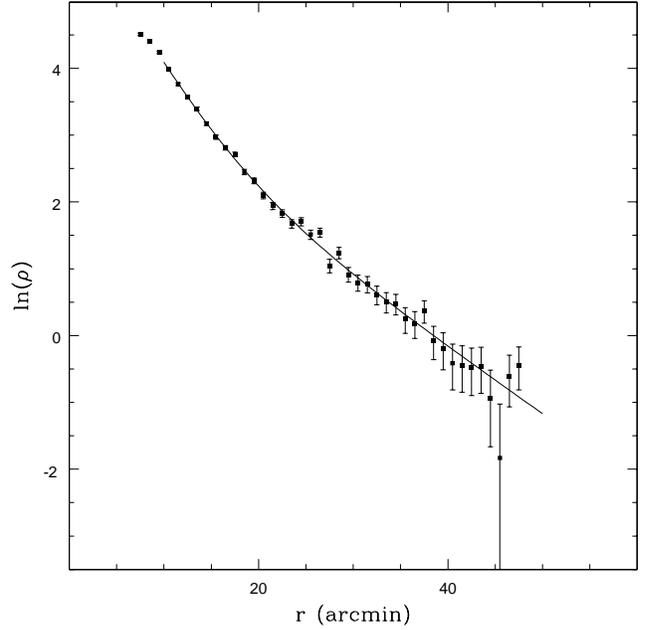}
   \caption{Two exponential-fit, in the range 
10$' < r < 50'$ gives a slightly better fit than the S\'ersic law.
}
              \label{Fig 2 expo.}
    \end{figure}

The decomposition of the profile into two exponentials suggests that
we are observing in the spheroid profile the contribution of a bulge,
of scale length of 3.85$'$ (560 pc), and an extended spheroid 
of lower density and much larger scale length (1450 pc).
The individual profiles and the fractional contributions of these 
two components are presented in Figure 14. 

   \begin{figure}
   \centering
\includegraphics[ width=9cm]{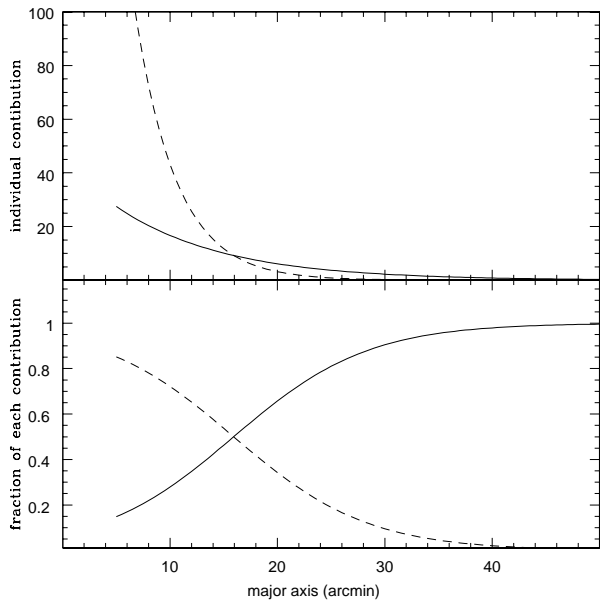}
   \caption{The profiles of the two exponentials of equation (2) are
shown in the top panel. In the lower panel we trace the contribution
of each to the total count. 
}
              \label{Fig 2exponentials}
    \end{figure}
 
\subsection{The extent of the intermediate-age population}

From their identification of C stars, Letarte et al. (2002) have shown 
that the stellar population of NGC 6822 extends to a surprisingly large
distance, when compared to the bright central core of the galaxy. The
CFHT12k mosaic reaches only 20$'$ from the centre, leaving a substantial
area of the ellipse of Fig. 11 not surveyed. The SDSS colours cannot
unambiguously pick up C stars since, as we have shown in Sec. 3.3, the
pollution by background galaxies is severe.
However, we now have a reliable estimate of the contribution of
interlopers thanks to the two Megacam fields obtained to evaluate the
properties of the foreground/background. 
This permits us to correct the C star counts in the region
surrounding NGC 6822 and obtain their density profile.

Figure 15 presents the surface density of C stars, as defined from the
SDSS colours. The profile is obtained by counting C star candidates in elliptical
annuli having the proper PA and ellipticity and subtracting the background density.
The profile reveals that intermediated-age stars belong to the bulge as well as
to the outer spheroid. The small numbers involved prevent us to rule out the 
possibility that a few of our C stars may belong to the HI disk component.
According to the profile in Fig. 15, C stars are detectable in appreciable 
number only up to $\approx 40'$ but the recent discovery of a globular cluster
at $\approx 79'$ (Hwang et al. 2005) shows that NGC 6822 extends beyond 12 kpc.
In recent years evidences on the existence of such extended structures around
dwarf galaxies have accumulated, for example: Leo A (Vansevi\v cious et al. 
2004), LMC (Minniti et al. 2003), IC 10 (Demers et al., 2004), 
DDO 187 (Aparicio et al. 2000).
 Once more, when securely identified, these
stars will provide valuable kinematic probes to better understand
the peculiar dynamics of NGC 6822. 
   \begin{figure}
   \centering
\includegraphics[ width=9cm]{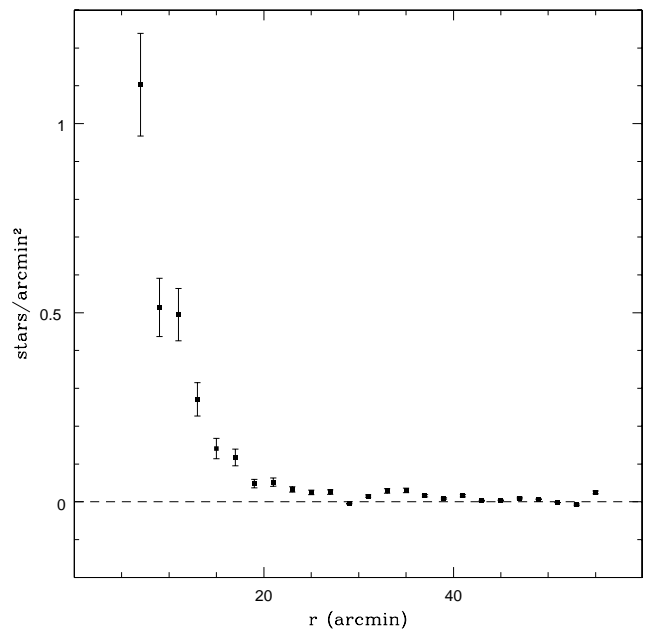}
   \caption{The surface density profile of C stars, corrected for
the background pollution but not corrected for the gaps between CCDs.
}
              \label{Fig. C profile}
    \end{figure}

\section{Discussion}

The distribution of the RGBs, displayed in Fig. 9, reveals the presence of
elliptically shaped spheroid. The spatial distribution of these old stars
contrasts with the youngest population, restricted to the central parts where
the HI is observed. To underline the difference between the two spatial distributions,
we present in  Figure 16, on the same scale,
 the spatial distribution of
some 30,000 main sequence stars, selected as stars with  $(g'-r')_0 < 0.0$.
The blue stars have also an elliptical distribution but the orientation of the major
axis of their ellipse follows the HI disk and is nearly at right angles to the major
axis of the spheroid. One notices some stars with similar colours distributed
all over the field. These are stars with spurious colours whose photometry
is affected by the presence of extremely bright stars. They are seen as
``clusters'' which are located near bright stars. The brightest and most 
obvious one is in the lower right of the plot. 

   \begin{figure}
   \centering
\includegraphics[ width=9cm]{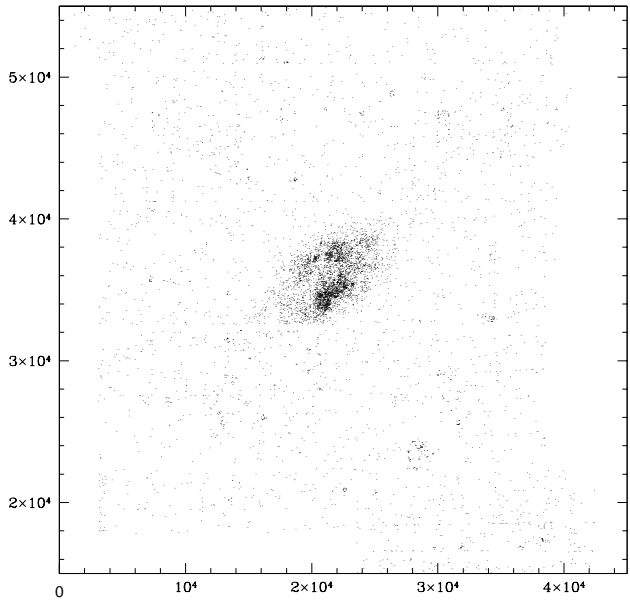}
   \caption{Spatial distribution of main sequence stars earlier than
A0. This figure is to be compared to Fig. 9.
}
              \label{Fig. blue map}
    \end{figure}

The profile of the spheroid which can be fitted to two exponentials
is not unique among dwarf galaxies. Indeed,
Sandage \& Binggeli (1984) observed that the surface brightness profile
of some dE galaxies of the Virgo Cluster could not be fitted with a
single exponential component. Binggeli \& Cameron (1991) matched an
exponential profile to the surface brightness of the outermost parts
of these galaxies but could not fit all radii,
indicating that these objects have at least two components.
They call these galaxies dwarf lenticulars (dS0s). It has later been
found that some of the dS0s in Virgo have a disk-like structure due to
the presence of bars and/or spiral patterns (Barazza et al. 2002).
More recently, from an analysis of the structural parameters
of dwarf galaxies in
the Coma cluster, Aguerri et al. (2005) identified two types of
dwarfs: the dEs with a S\'ersic profile and the dS0s with a S\'ersic +
exponential profile. They note that the dS0 galaxies have a mean ellipticity 
of 0.36 $\pm$ 0.05, flatter than dEs and essentially similar to what we 
find for NGC 6822. They suggest that  harassment by nearby spirals could
explain the origin of dS0s.
Closer to the Local Group, the dSO galaxy NGC 59 in the Sculptor Group, may 
also have a polar ring. The HI observations by Beaulieu et al. (2006) do not 
have sufficient resolution to confirm this feature.
Two structural features of NGC 6822, the central bar and the two
exponential profile, imply that its spheroid could be identified
as a dSB0 galaxy. Furthermore, the presence of the HI disk nearly at right 
angles to the major axis of the spheroid reveals that we are seeing a 
typical polar ring galaxy (hereafter, PRG) configuration. PRGs are indeed 
composed of an early-type host galaxy surrounded by a gaseous and stellar 
perpendicular ring or disk with characteristic typical of late-type galaxies.
This hypothesis  has been confirmed by our radial velocity  measurements of 
nearly 100 C stars which revealed that the spheroid is rotating around its minor
axis, thus the spheroid and the HI disk angular momenta are nearly at right 
angles (Demers et al., 2006).

The size and mass of the polar ring is of little help to assess whether
we are observing a genuine PRG. Indeed, both models and observations 
show that PRGs may have various radii and masses (see e.g. Bournaud \& Combes, 
2003).
 The inclination of the polar ring found in NGC 6822 ($\approx 25^\circ
$) is large (most PRGs are inclined by less than this value) but some PRGs
are observed with much higher inclinations (e.g. the ring of NGC 660 is at 
$\approx 45^\circ$).

The first major study of the class of PRGs was done by Whitmore et al.
(1990). They presented data of some 100 galaxies but at that time only a
few were kinematically confirmed. The number of kinematically confirmed
PRGs is rising slowly, less than two dozens are presently known. The two
nearest ones being NGC 660, at 13 Mpc, and NGC 2685 at 13.5 Mpc. In this
context, the identification of NGC 6822 as a PRG offers a great leap forward
to better understand this phenomenon.

Some caution is however due since the two most likely mechanisms for the
formation of PRGs, namely merging and tidal accretion, both imply the 
interaction with some neighbour while NGC 6822 is, {\it presently}, an 
isolated galaxy. The tiny NW HI cloud suggested by Komiyama et al. (2003) 
to be responsible of the overall SFR enhancement observed in NGC 6822 (see e.g. 
Gallart et al., 1996) and of the SE bent of the HI density contours is 
certainly too small to explain the present configuration of NGC 6822. 
The Milky Way could have influenced, in the past, the morphology of NGC 6822. 
These galaxies are currently moving apart at 44 km s$^{-1}$ at this rate
they were 50\% 
closer a few Gyr ago. More exotic formation mechanisms,
which do not require two galaxy encounters, have also been proposed (see
Combes 2005 for a review) .

PRGs investigated by Ritcher et al. (1994) were found to be 
much more luminous in the far infrared (FIR) than typical early-type
galaxies (S0). But, the few kinematically confirmed PRGs did not show
such excess. Taking for NGC 6822 L$_{FIR}$ = 116 Jy and
L$_B$ = 1.88 Jy, from the NED/IPAC database,  
we calculate a log(L$_{FIR}$/L$_B$) = 1.79 making NGC 6822 a galaxy with
an extreme ratio, when compared to the Ritcher et al. sample. 
Furthermore, the HI mass of NGC 6822 is quite normal for a dIrr of its 
luminosity implying that the HI seen as a disk may be intrinsic to NGC 6822. 
Its very red (60$\mu$/100$\mu$) infrared colour is not particularly high
suggesting that the dust, present in the central region, is heated
by the blue stars present in abundance and not by numerous red giants.

Another feature often found among PRGs is that the HI rotates faster than 
predicted for normal spiral galaxy Tully-Fisher (TF) relation (see Iodice et al. 
2003) while, as Combes (2005) has shown, on a sample of five well 
studied PRGs, the host galaxy rotation is in fair agreement with the TF relation.
All this has been interpreted as evidence of a DM halo flattened along 
the HI disk. The rotation along the entire length of the major axis of the
spheroid of NGC 6822 has not yet been mapped. For the inner 15 arcmin,
Demers et al. (2006) found that the spheroid rotates, within one 
standard deviation, following the red TF relation as given by Giovannelli 
et al. (1997). The HI disk appears also to follow the TF relation, using the
rotation curve from Weldrake et al. (2003) and taking M$_I=-16.1$ for the
absolute magnitude (NED/IPAC).

\section{Summary}
We have determined the reddening map over an area of $\approx 6$ square degrees
in the direction of NGC 6822. This reddening map is used to deredden the 
colour-magnitude and colour-colour diagrams. The surface density profile of
stars selected in the RGB region of the CMD has revealed the existence of a 
stellar spheroid that we could trace up to $36 '$ from the centre of NGC 6822.
This spheroid is of elliptical shape with a major axis at position angle of
65$^\circ$, thus nearly at right angles with the major axis of the HI disk 
described by de Blok \& Walter (2000). Such configuration suggests that NGC 6822
is a polar ring galaxy.  The stellar density profile is well fitted by a 
two-exponential law with scale lengths of 10$'$ and $3.9'$. We confirm that the
spheroid contains, at least up to 40$'$, 
a detectable intermediate-age
population revealed by the presence of C stars.

The identification of confirmed members at the extremities
of its major axis is crucial to our understanding of the dynamical
evolution of NGC 6822. Indeed, the rotation curve of the spheroid 
must be mapped
to the limit in order to assess its full mass and see if a
turn over, revealing past tidal interactions is present. For these
reasons, our search for C stars in the periphery continues.

The proximity of NGC 6822  offers a unique opportunity to investigate
scenarios leading to the formation of the polar disk. One could,
in principle, determine which of the stellar populations belongs to each
dynamical system thus permitting to date the event and identify
the nature of the intruder.

\begin{acknowledgements}
This research
is funded in parts (S. D.) by the Natural Sciences and Engineering Research
Council of Canada. We are grateful to Yannick Mellier and the Terapix
to have promptly analysed our Megacam data.
\end{acknowledgements}

\end{document}